\documentclass{ifacconf}
\usepackage{graphicx}      %
\usepackage{natbib}        %
\usepackage{tikz}
\usepackage{amsmath}
\usepackage{amssymb}
\usepackage{amsfonts}

\usepackage{algorithm,algorithmic}

\usepackage{caption}
\usepackage{subcaption}

\usepackage{mathtools}

\usepackage{enumitem}

\newcommand{\yref}{y^{\text{ref}}_t}

\begin{document}

\begin{frontmatter}

\title{Observer-Feedback-Feedforward Controller Structures in Reinforcement Learning\thanksref{footnoteinfo}}

\thanks[footnoteinfo]{
This research was financially supported by the project \emph{NewLEADS - New Directions in Learning Dynamical Systems} (contract number: 621-2016-06079), funded by the Swedish Research Council.}

\author[First]{Ruoqi Zhang}
\author[First]{Per Mattson}
\author[First]{Torbj\"{o}rn Wigren}

\address[First]{Department of Information Technology, Uppsala University,
   SE-75105 Uppsala, SWEDEN (e-mail: firstname.lastname@it.uu.se).}

\begin{abstract}                %
The paper proposes the use of structured neural networks for reinforcement learning based nonlinear adaptive control. The focus is on partially observable systems, with separate neural networks for the state and feedforward observer and the state feedback and feedforward controller.
The observer dynamics are modelled by recurrent neural networks while a standard network is used for the controller. As discussed in the paper, this leads to a separation of the observer dynamics to the recurrent neural network part, and the state feedback to the feedback and feedforward network. The structured approach reduces the computational complexity and gives the reinforcement learning based controller an {\em understandable} structure as compared to when one single neural network is used. As shown by simulation the proposed structure has the additional and main advantage that the training becomes significantly faster. 
Two ways to include feedforward structure are presented, one related to state feedback control and one related to classical feedforward control. The latter method introduces further structure with a separate recurrent neural network that processes only the measured disturbance. When evaluated with simulation on a nonlinear cascaded double tank process, the method with most structure performs the best, with excellent feedforward disturbance rejection gains.   
\end{abstract}


\end{frontmatter}
\section{Introduction}

The paper proposes new structured architectures
for reinforcement learning (RL) when solving partially observable control problems. 
The first contribution is depicted in Fig.~\ref{fig:obserferff-network-1}, in which the single neural network (NN) that is routinely used to solve nonlinear control problems \citep{heess2015memory-based,hausknecht2015dqn_rnn_modelfree, ha2018recurrent_modelbased,  zhang2019solar, meng2021memory} is split into separate parts with an observer that estimates a low-dimensional state, and a state feedback controller. Furthermore, a critic is included for training purposes. 

The rationale behind the introduction of a prior structure is to reduce the complexity of the network, which gives a number of advantages. With this structure, the number of unknowns is reduced, and the controller uses a low-dimensional state representation. This will, as discussed throughout the paper, lead to quicker training and increased controller performance, using less computations. 
In addition, the widely discussed problem of poor {\em understandability} of RL-based technical solutions in general is addressed. Well-known techniques for the analysis of signals 
 in conventional control structures can be applied, like separating the effect of feedforward and feedback, thereby e.g. focusing the control objective contributions to the relevant frequency bands.

\usetikzlibrary{calc, fit, positioning}
\def\layersep{0.91}
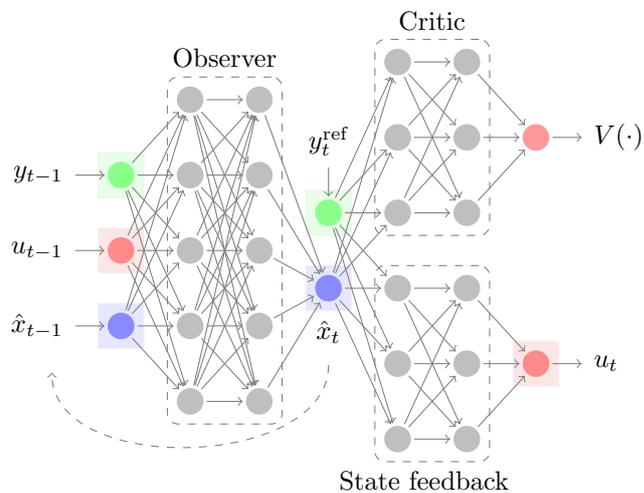
\begin{figure}[htbp!]
    \centering
    \begin{tikzpicture}[shorten >=1pt,->,draw=black!50, 
    node distance = \layersep,
every pin edge/.style = {<-,shorten <=1pt},
        neuron/.style = {circle,fill=black!25, minimum size=10 pt,inner sep=0pt},
     no color neuron/.style = {circle, fill=black!10, minimum size=10 pt,},
     dummy neuron/.style = {circle, minimum size=10 pt,},
  input neuron/.style = {neuron, fill=green!40},
 output neuron/.style = {neuron, fill=red!40},
 hidden neuron/.style = {neuron, fill=blue!40},
         annot/.style = {text width=4em, text centered},
    shorten >=1pt, shorten <=1pt,
    legend/.style={font=\large\bfseries},
  ]

    \node[input neuron, pin=left: $y_{t-1}$ ] (I-1) at (0, -2){};
    \node [fit=(I-1), draw, fill=green, opacity=0.1] (observation box) {};
    \node[output neuron, pin=left: $u_{t-1}$ ] (I-2) at (0, -3){};
    \node [fit=(I-2), draw, fill=red, opacity=0.1] (action box) {};
    \node[hidden neuron, pin=left: $\hat{x}_{t-1}$ ] (I-3) at (0, -4){};
    \node [fit=(I-3), draw, fill=blue, opacity=0.1] (hidden box1) {};

    \foreach \name / \y in {1,...,5}
        \path[yshift=0cm]
            node[neuron] (O1-\name) at (\layersep, -\y cm) {};
    \foreach \source in {1,...,3}
        \foreach \dest in {1,...,5}
            \path (I-\source) edge (O1-\dest);
    \foreach \name / \y in {1,...,5}
        \path[yshift=0.cm]
            node[neuron] (O2-\name) at (2*\layersep  cm, -\y cm) {};
    \foreach \source in {1,...,5}
        \foreach \dest in {1,...,5}
            \path (O1-\source) edge (O2-\dest);   

    \node [fit=(O1-1) (O2-5), draw, dashed, rounded corners, label={[rotate=0]above:Observer}] (hidden box1) {};

    \node[input neuron, pin=above: $\yref$] (O3-1) at (3*\layersep, -2.5) {};
    \node [ fit=(O3-1), draw, fill=green, opacity=0.1] (reference signal box) {};
    \node[hidden neuron] (O3-2) at (3*\layersep, -3.5) {};
    \node [label=below: $\hat{x}_{t}$, fit=(O3-2), draw, fill=blue, opacity=0.1] (hidden box2) {};
    \node [dummy neuron] (Dummy-2) at (3*\layersep, -4.3) {};
   \node [dummy neuron] (Dummy-1) at (-\layersep, -4.4) {};
    \path [dashed] (Dummy-2) [out=-90, in=-90] edge (Dummy-1);

    \foreach \source in {1,...,5}
        \foreach \dest in {2}
            \path (O2-\source) edge (O3-\dest);
            
    \foreach \name / \y in {4,...,6}
        \path[yshift=0.5cm]
            node[neuron] (A1-\name) at (4*\layersep, -\y cm) {};
    \foreach \source in {1,...,2}
        \foreach \dest in {4,...,6}
        \path (O3-\source) edge (A1-\dest);
        
    \foreach \name / \y in {4,...,6}
        \path[yshift=0.5cm]
            node[neuron] (A2-\name) at (5*\layersep, -\y cm) {};
    \foreach \source in {4,...,6}
        \foreach \dest in {4,...,6}
        \path (A1-\source) edge (A2-\dest);
        
   \node [fit=(A1-4) (A2-6), draw, dashed, rounded corners, label={[rotate=0, xshift = -3]below:State feedback}] (Actor box) {};
        
    \node[output neuron,pin={[pin edge={->}]right:$u_t$}] (OA) at (6*\layersep, -4.5 cm) {};
    \node [fit=(OA), draw, fill=red, opacity=0.1] (action box) {};

    \foreach \source in {4,...,6}
        \path (A2-\source) edge (OA);
    
    \foreach \name / \y in {1,...,3}
        \path[yshift=0.5cm]
            node[neuron] (C1-\name) at (4*\layersep, -\y cm) {};
    \foreach \source in {1,...,2}
        \foreach \dest in {1,...,3}
        \path (O3-\source) edge (C1-\dest);
        
    \foreach \name / \y in {1,...,3}
        \path[yshift=0.5cm]
            node[neuron] (C2-\name) at (5*\layersep, -\y cm) {};

    \foreach \source in {1,...,3}
        \foreach \dest in {1,...,3}
        \path (C1-\source) edge (C2-\dest);
        
    \node[output neuron, pin={[pin edge={->}]right:$V(\cdot)$}] (V) at (6*\layersep, -1.5 cm) {};
    \foreach \source in {1,...,3}
        \path (C2-\source) edge (V);
        
    \node [fit=(C1-1) (C2-3), draw, dashed, rounded corners, label={[rotate=0]above:Critic}] (Critic Box) {};
    \end{tikzpicture}
    \caption{Structured neural network model.}
    \label{fig:obserferff-network-1}
\end{figure}

The proposed training procedure is based on the recurrent model-free RL in a history-based way \citep{heess2015memory-based,meng2021memory, hausknecht2015dqn_rnn_modelfree}.
    Compared to other algorithms attempting to first infer the state, known as a belief state  \citep{kaelbling1998planning}, and then apply standard RL methods to the belief state, the model-free history-based methods do not need an explicit model of the dynamic system. Instead, they use the measured signals directly.

The second contribution 
examines if the structure of Fig.~\ref{fig:obserferff-network-1} is general enough for nonlinear control. By the selection of a recurrent neural network (RNN), see e.g. \citet{elman1990finding}, for the observer and a standard forward neural network for the state feedback controller, 
arguments related to the separation theorem, see e.g. \cite{soderstrom2002stochastic_system}, together with simulations show why the observer dynamics and state feedback can be correctly separated in the case of a linear system. The third contribution builds on the work of  \citet{sternad1988lqg} where (linear) optimality is shown for the case where the dynamics of the {\em measurable} disturbance is separately observed.
 The paper therefore proposes a similar structure, with a separate recurrent neural network that only receives input from the disturbance. The paper then shows that the tentative advantages do also materialize for nonlinear systems by simulation experiments on a cascaded tank laboratory process.  
 While this laboratory process is relatively simple to control using standard control methods, 
the paper still motivates the use of controller structure for the solution of advanced nonlinear control problems with reinforcement learning techniques.

In previous related work by \citet{heess2015memory-based} the whole history is used as the input to the RL algorithms, using
RNNs as a way to summarize it. In this paper, it is rather shown by simulation of a linear system that the internal state of the RNNs is a valid estimate of a true state representation. This suggests viewing the RNN as an observer and therefore also adjusting the output size of the RNN to an (upper bound) of the size of a minimal state representation and letting the state feedback controller learn to control from this smaller state representation. 

The paper is organized with the problem statement and background in Section~\ref{sec::ps}. 
The structured networks are introduced in Section~\ref{sec::model} while performance is addressed in Section~\ref{sec::vali:observer} and \ref{sec::vali:feedforward}. Conclusions end the paper in Section~\ref{sec::conclusion}.

\section{Problem Statement and Background}
\label{sec::ps}

Consider a discrete-time nonlinear dynamical system described by the state-space model, 
\begin{align}
	\label{eq:ps}
	\begin{split}
		x_{t+1} &= f(x_t, u_t, w_t, d_t)\\
		y_{t} &= h(x_t, v_t)
	\end{split}
\end{align}
where $f$ and $h$ are unknown nonlinear functions, 
$x_t \in \mathbb{R}^{n_x}$ is the state vector of the dynamic system, $u_t \in \mathbb{R}^{n_u}$ is the input vector, and $w_t \in \mathbb{R}^{n_w}$ is the process noise. The measured output vector $y_t \in \mathbb{R}^{n_y}$ is corrupted by the noise $v_t \in \mathbb{R}^{n_v}$, and $d_t \in \mathbb{R}^{n_d}$ is the \emph{measurable} disturbance vector.

The task considered in the paper is to learn a controller that uses the measured outputs $y_t$ and disturbance $d_t$ to track a reference signal $\yref$. Specifically, the aim is to see how RL-based methods can be improved by enforcing control structures often used in control theory. 

Some background on RL is given in Section~\ref{sec:backgroundRL}, while common control structures used in linear control are discussed in Section~\ref{sec:lin}.

\subsection{Background on RL}\label{sec:backgroundRL}
A fully observable discrete-time dynamic system can be modelled as a Markov Decision Process (MDP) consisting of a set of states $\mathcal{X}$, a set of the inputs (actions) $\mathcal{U}$, the reward function $\mathcal{R}\in \mathbb{R}$ and the discount factor $\gamma \in (0,1]$. In a fully observed MDP it is assumed that noise-free measurements of $x_t$ are available. 
The tracking problem with different reference signal values can be solved by multi-goal RL \citep{plappert2018multigoal}, where the reward function not only depends on the state $x_t$ and input $u_t$ but also depends on the reference signal $\yref$, i.e. $r_t = \mathcal{R}(x_t, u_t, \yref)$. The state-value function
estimates how rewarding the current state is in terms of expected future returns. The universal value function \citep{schaul2015universal-value-functions} which depends not only on the state but also on the reference signal $\yref$ under a controller $\pi$ is defined as 
\begin{equation}\label{eq:V}
	V_\pi(x_t, \yref) = \mathbb{E}_\pi \left[G_t \vert x_t, \yref\right] = \mathbb{E}_\pi  \left[\sum_{k=0}^\infty \gamma^{k}r_{k+t} \right].
 \end{equation}
where $\mathbb{E}_\pi[\cdot]$ denotes the expected value while using the controller $\pi$ and $t$ is any time step. 
The goal is then to find a controller $\pi$ that maximizes \eqref{eq:V}.
To achieve this, both the controller $\pi$ and an estimate of the value function $V$, called a \textit{critic}, are parameterized by neural networks and learned from experience. 
\begin{rem}
The value function in \eqref{eq:V} depends on the rewards $r_t$. In this contribution, the goal is that the output $y_t$ should follow the reference $\yref$, so the rewards 
\[
r_t = -\|y_t - \yref\|^2_2
\]
are used. That is, to maximize the reward, $y_t$ should be as close to $\yref$ as possible in the least squares sense.
\end{rem}

While standard methods in RL often assume noise-free measurements of the full state $x_t$ in \eqref{eq:ps}, the case considered in this paper is when only the outputs $y_t$ and disturbances $d_t$ are measured. 
In RL this setting is called a Partially Observable Markov Decision Process (POMDP).
Without access to the full state, one possibility is to consider the full history of measurements to be an alternative state representation. That is, the state is 
\begin{equation}\label{eq:hist}
\varphi_t =\left\{ (y_k, u_k, d_k) \right\}_{k=1}^t.
\end{equation}
However, since $\varphi_t$ grows with $t$ this is not a suitable choice.
In practice, therefore 
\begin{equation}
\tilde{\varphi}_t =\left\{ (y_k, u_k, d_k) \right\}_{k=t-N}^t
\end{equation}
for some fixed $N$ is often assumed to capture the full state, see for example \citep{mnih2015dqn} where $N=4$ past observations are used as a surrogate for the full state.
However, this assumes that the state can indeed be captured by a few of the past measurements.

\citet{heess2015memory-based} instead use
an approach where RNNs compress the full history in~\eqref{eq:hist}. This is made possible by the fact that contrary to standard static networks, an RNN can store an internal state between time steps.
In  Section~\ref{sec::model} a similar idea is used to design a structured approach to explicitly learn an observer that estimates the state using standard RL methods.

 \subsection{The linear case}\label{sec:lin}
 When new methods are developed, it is a good idea to make sure that they are reasonable for simple cases, like when \eqref{eq:ps} is a linear dynamic system. In the linear setting, the following assumptions are common:
\begin{enumerate}[label=A\arabic*)]
	\item 
	\label{enumerate:linearsys}
	the system is given by 
 \begin{equation}
	\begin{aligned}
		x_{t+1} &= A x_t + B u_t + N\begin{bmatrix} w_t \\ d_t \end{bmatrix} \\
		y_t &= C x_t + v_t, \\
		\label{eq:linearsys}
	\end{aligned}
 \end{equation}
	where  $A, B, N$ and $C$ are matrices. 

	\item 
	\label{enumerate:stable}
	the system is stabilizable and detectable.
	\item  \label{enumerate:measurable}
	$d_t$ is measurable.
	\item
	\label{enumerate:gaussian}
	$w_t$ and $v_t$ are zero-mean Gaussian random processes.
	\item 
	\label{enumerate:criterion}
	the performance criterion for the determination of the feedback controller is 
	\begin{align}
		\label{eq:criterion}
		J = \mathbb{E}\left[y_t^TQ_1y_t + u_t^TQ_2u_t\right].
	\end{align} 
\end{enumerate}
\begin{rem}
The criterion that will be used by the RL-methods below is that of \eqref{eq:V}. While this is not exactly the same as \eqref{eq:criterion}, the optimal controllers have similar properties.
\end{rem}

It is then proved in textbooks on optimal control, e.g. by \citet{soderstrom2002stochastic_system},  that the optimal state feedback controller is separated into a Kalman filter observer and an optimal state feedback controller, with states replaced by the estimates of the Kalman filter. The extension to include optimal feedforward control is derived by \cite{sternad1988lqg}. In the linear case, the result is an additional additive optimal feedforward controller. 

It can be noted that while the network in Fig.~\ref{fig:obserferff-network-1} 
gives a general nonlinear controller, its structure still is consistent with the optimal structure for a linear state space LQG controller.
For nonlinear systems the separation theorem does not hold in general, so the optimal state feedback controller and the optimal observer can no longer be designed separately. To compensate for this the proposed method in Section~\ref{sec::model} trains the observer and controller at the same time, end-to-end. 

Together, this motivates a structured combination of an RNN for observation and a standard neural network for state feedback control based on estimated states as in Fig.~\ref{fig:obserferff-network-1}.

\section{Observer-Feedback-Feedforward Structures for RL}
\label{sec::model}
\subsection{Low order observer}

\citet{heess2015memory-based} viewed the full history \eqref{eq:hist} as the true state of the system, and the RNNs were used to compress the history. Since both the critic and controller need to know the state, two fully separate RNNs were used to learn these. Below this setting is referred to as ``unstructured''.

However, this state representation has by design a much higher dimension than a minimal state representation, which in turn makes it harder to learn a good 
controller from this state. 
This, together with the discussion in Section~\ref{sec:lin}, motivates the use of a structured network as in Fig.~\ref{fig:obserferff-network-1} divided into a dynamic observer using an RNN that estimates a relatively \emph{low-order} representation of the state and a static state feedback controller. 
At each time step~$t$, the observer receives measurements of $y_{t-1}$ and $d_{t-1}$. 
This information is then combined with the last input $u_{t-1}$ and the last state estimate $\hat{x}_{t-1}$ to update the current estimate of the state of the system. 
This estimated state can then be used both in the critic and for state feedback control, with much improved performance as discussed in Sections~\ref{sec::vali:observer} and~\ref{sec::vali:feedforward}.

\usetikzlibrary{calc, fit, positioning}
\def\layersep{0.91}

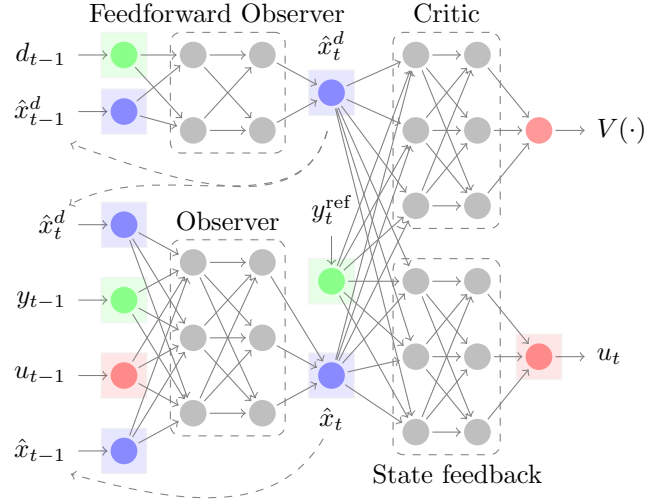
\begin{figure}[htbp!]
    \centering
    \begin{tikzpicture}[shorten >=1pt,->,draw=black!50, 
    node distance = \layersep,
every pin edge/.style = {<-,shorten <=1pt},
        neuron/.style = {circle,fill=black!25, minimum size=10 pt,inner sep=0pt},
     no color neuron/.style = {circle, fill=black!10, minimum size=10 pt,},
     dummy neuron/.style = {circle, minimum size=10 pt,},
  input neuron/.style = {neuron, fill=green!40},
 output neuron/.style = {neuron, fill=red!40},
 hidden neuron/.style = {neuron, fill=blue!40},
         annot/.style = {text width=4em, text centered},
    shorten >=1pt, shorten <=1pt,
    legend/.style={font=\large\bfseries},
  ]
    \node[input neuron, pin=left: $d_{t-1}$ ] (D-1) at (0, -0.5){};
	\node [ fit=(D-1), draw, fill=green, opacity=0.1] (disturbance box2)  {};
	\node[hidden neuron, pin=left: $\hat{x}_{t-1}^d$ ] (D-2) at (0, -1.25) {};
	\node [ fit=(D-2), draw, fill=blue, opacity=0.1] (disturbance box2)  {};
	
    \foreach \name / \y in {1,...,2}
        \path[yshift=0.5cm]
            node[neuron] (OF1-\name) at (\layersep, -\y cm) {};
    \foreach \source in {1,...,2}
        \foreach \dest in {1,...,2}
            \path (D-\source) edge (OF1-\dest);
    \foreach \name / \y in {1,...,2}
        \path[yshift=0.5cm]
            node[neuron] (OF2-\name) at (2*\layersep  cm, -\y cm) {};
    \foreach \source in {1,...,2}
        \foreach \dest in {1,...,2}
            \path (OF1-\source) edge (OF2-\dest);   

    \node [fit=(OF1-1) (OF2-2), draw, dashed, rounded corners, label={[rotate=0, xshift = -4]above:Feedforward Observer}] (hidden box1) {};
    
	\node[hidden neuron] (XD-t) at (3*\layersep, -1) {};
    \node [label=above: $\hat{x}_{t}^d$, fit=(XD-t), draw, fill=blue, opacity=0.1] (hidden disturbance) {};
    \foreach \source in {1,...,2}
        \path (OF2-\source) edge (XD-t);

    \node[hidden neuron, pin=left: $\hat{x}_{t}^d$ ] (XD-1) at (0, -2.75){};
    \node [fit=(XD-1), draw, fill=blue, opacity=0.1] (observation box) {};
    
    \node [dummy neuron] (Dummy-2) at (3*\layersep, -1.3) {};
    \node [dummy neuron] (Dummy-1) at (-\layersep, -2.6) {};
    \node [dummy neuron] (Dummy-3) at (-\layersep, -1.6) {};
    \path [dashed] (Dummy-2) [out=-105, in=30] edge (Dummy-1);
    \path [dashed] (Dummy-2) [out=-105, in=-20] edge (Dummy-3);
    
    \node[input neuron, pin=left: $y_{t-1}$ ] (I-1) at (0, -3.75){};
    \node [fit=(I-1), draw, fill=green, opacity=0.1] (observation box) {};
    \node[output neuron, pin=left: $u_{t-1}$ ] (I-2) at (0, -4.75){};
    \node [fit=(I-2), draw, fill=red, opacity=0.1] (action box) {};
    \node[hidden neuron, pin=left: $\hat{x}_{t-1}$ ] (I-3) at (0, -5.75){};
    \node [fit=(I-3), draw, fill=blue, opacity=0.1] (hidden box1) {};

    \foreach \name / \y in {1,...,3}
        \path[yshift=-2.25cm]
            node[neuron] (O1-\name) at (\layersep, -\y cm) {};
    \foreach \source in {1,...,3}
        \foreach \dest in {1,...,3}
            \path (I-\source) edge (O1-\dest);
    \foreach \dest in {1,...,3}
        \path (XD-1) edge (O1-\dest);
    
    \foreach \name / \y in {1,...,3}
        \path[yshift=-2.25cm]
            node[neuron] (O2-\name) at (2*\layersep  cm, -\y cm) {};
    \foreach \source in {1,...,3}
        \foreach \dest in {1,...,3}
            \path (O1-\source) edge (O2-\dest);   

    \node [fit=(O1-1) (O2-3), draw, dashed, rounded corners, label={[rotate=0]above:Observer}] (hidden box1) {};

    \node[input neuron, pin=above: $\yref$] (O3-1) at (3*\layersep, -3.5) {};
    \node [ fit=(O3-1), draw, fill=green, opacity=0.1] (reference signal box) {};
    \node[hidden neuron] (O3-2) at (3*\layersep, -4.75) {};
    \node [label=below: $\hat{x}_{t}$, fit=(O3-2), draw, fill=blue, opacity=0.1] (hidden box2) {};
    
    \node [dummy neuron] (Dummy-2) at (3*\layersep, -5.4) {};
    \node [dummy neuron] (Dummy-1) at (-\layersep, -6.0) {};
    \path [dashed] (Dummy-2) [out=-120, in=-15] edge (Dummy-1);

    \foreach \source in {1,...,3}
        \foreach \dest in {2}
            \path (O2-\source) edge (O3-\dest);
            
    \foreach \name / \y in {4,...,6}
        \path[yshift=0.5cm, xshift=0.2cm]
            node[neuron] (A1-\name) at (4*\layersep, -\y cm) {};
    \foreach \source in {1,...,2}
        \foreach \dest in {4,...,6}
        \path (O3-\source) edge (A1-\dest);
    \foreach \dest in {4,...,6}
        \path (XD-t) edge (A1-\dest);
        
    \foreach \name / \y in {4,...,6}
        \path[yshift=0.5cm, xshift=0.1cm]
            node[neuron] (A2-\name) at (5*\layersep, -\y cm) {};
    \foreach \source in {4,...,6}
        \foreach \dest in {4,...,6}
        \path (A1-\source) edge (A2-\dest);
        
   \node [fit=(A1-4) (A2-6), draw, dashed, rounded corners, label={[rotate=0, xshift = 4]below:State feedback}] (Actor box) {};
        
    \node[output neuron,pin={[pin edge={->}]right:$u_t$}] (OA) at (6*\layersep, -4.5 cm) {};
    \node [fit=(OA), draw, fill=red, opacity=0.1] (action box) {};

    \foreach \source in {4,...,6}
        \path (A2-\source) edge (OA);
    
    \foreach \name / \y in {1,...,3}
        \path[yshift=0.5cm, xshift=0.2cm]
            node[neuron] (C1-\name) at (4*\layersep, -\y cm) {};
    \foreach \source in {1,...,2}
        \foreach \dest in {1,...,3}
        \path (O3-\source) edge (C1-\dest);
    \foreach \dest in {1,...,3}
        \path (XD-t) edge (C1-\dest);
        
    \foreach \name / \y in {1,...,3}
        \path[yshift=0.5cm, xshift=0.1cm]
            node[neuron] (C2-\name) at (5*\layersep, -\y cm) {};

    \foreach \source in {1,...,3}
        \foreach \dest in {1,...,3}
        \path (C1-\source) edge (C2-\dest);
        
    \node[output neuron, pin={[pin edge={->}]right:$V(\cdot)$}] (V) at (6*\layersep, -1.5 cm) {};
    \foreach \source in {1,...,3}
        \path (C2-\source) edge (V);
        
    \node [fit=(C1-1) (C2-3), draw, dashed, rounded corners, label={[rotate=0]above:Critic}] (Critic Box) {}; 
    \end{tikzpicture}
    \caption{Network structure with observers for feedback and feedforward.}
   
    \label{fig:obserferff-network-2}
\end{figure}

\subsection{Separate feedforward observer}

In linear and in some structured non-linear state space based control systems, feedforward control is achieved by means of a state augmentation, see e.g. \cite{wigren2017loop}. Briefly, states modelling the feedforward dynamics are included, a fact that results in a joint observer for feedback and feedforward dynamics. The state controller is then obtained by criterion minimization, however {only the states generating the output signal}, i.e. that are directly related to the control objective, are extracted to appear in the performance metric.  This remaining separation into a state observer and a state controller in the case of feedforward control for linear systems, motivates the use of the structure of Fig.~\ref{fig:obserferff-network-1} also for nonlinear feedback and feedforward control, using end-to-end training.

However, further structuring can be motivated. In the linear case, the state controller consists of two terms, one term combining observed states generated from the feedforward observer and one term combining observed feedback states. Hence it is proposed to split {\em the observer of} Fig.~\ref{fig:obserferff-network-1} \emph{into two}, with one RNN  estimating states related to the disturbance feedforward dynamics, and another RNN estimating the states of the system. The NN corresponding to the state feedback controller is however not split into two, to retain nonlinear coupling effects. This structure can also be motivated by noting that an external disturbance $d_t$ is not affected by the systems state or inputs, but the state is affected by the disturbance. 
The combined state is finally used by the critic and the controller.  

As will be seen in Section~\ref{sec::vali:feedforward}, the structure of Fig.~\ref{fig:obserferff-network-2} performs significantly better than the alternative of Fig.~\ref{fig:obserferff-network-1}, most likely since it uses a more restrictive structure that forces the RL methods to learn a controller with a suitable structure. This typically leads to quicker convergence and more accurate results. 
The new NN-structures with the observers together with joint state feedback and feedforward control can be trained end-to-end with any standard RL algorithm. The Proximal Policy Optimization (PPO)  by \citet{schulman2017proximal} is used in this paper. 
It should be noted that the method is a variant of direct adaptive optimal control, see e.g. \citet{aastrom2013adaptive_control},
 meaning that the algorithm does not identify the dynamics of the system but directly learns the controller.

\section{Structured Observer-Controller Performance}
\label{sec::vali:observer}

Here simulation is used to investigate whether the observer learns (estimates) the true state of the system when trained by the end-to-end RL algorithm applied to the structure of Fig.~\ref{fig:obserferff-network-1}.

To address this question a linear system on the form \eqref{eq:linearsys} is considered. It is well-known that under mild conditions one state representation of a linear system can be changed to any other state representation by a linear transformation. This also makes it possible to determine if the RNN indeed generates a state consistent with~\eqref{eq:linearsys} when trained.

A linear system with 10 states was obtained by the Matlab function \texttt{drss}, which randomly generates a stable discrete-time linear system. The measurement noise $v_t$ and process disturbance $w_t$ were generated as zero-mean Gaussian random processes.  No measurable disturbance was present in this numerical experiment.

For training PPO was used on both a structured network as in Fig.~\ref{fig:obserferff-network-1} with separation of the observer and controller, and on a full RNN network without any structure as in \citep{heess2015memory-based}. In both cases, the aim is to minimize \eqref{eq:V} with the rewards $r_t = - (y_t - \yref)^2$.
To compare the setup with that discussed in Section~\ref{sec:lin}, it can be noted that conditions A1), A2) and A4) hold, while the feedforward related condition can be disregarded. The RL method also optimizes a criterion similar to that in A5). 

\begin{figure}[t]
    \centering
    \begin{subfigure}[b]{0.8\linewidth}
         \centering
         \includegraphics[width=\textwidth]{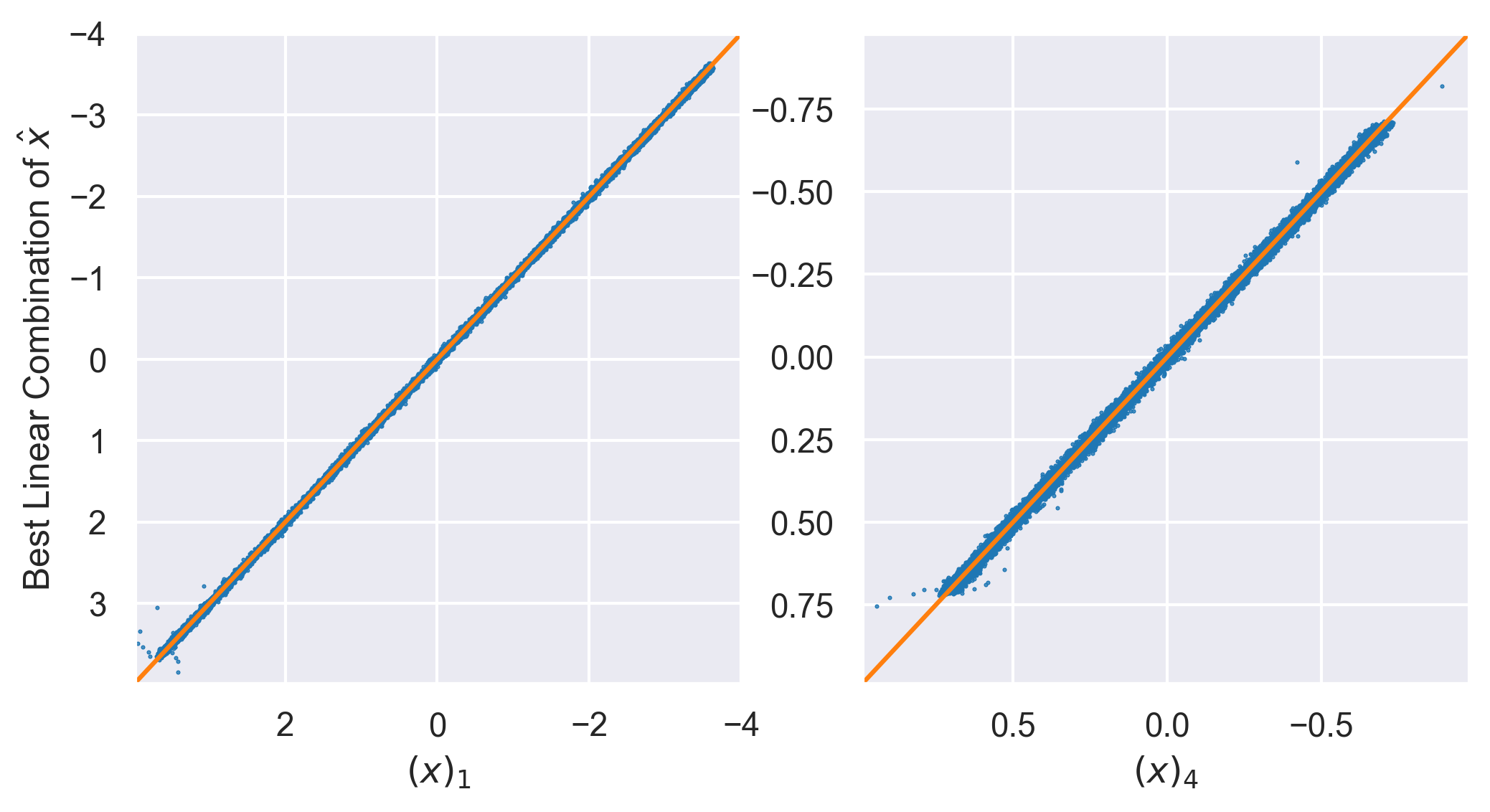}
         \caption{Correlation between the best linear combination of estimated state using the structure in Fig.~\ref{fig:obserferff-network-1} and the true states.}
         \label{fig:linear:correlation:roa}
     \end{subfigure}
     \hfill
     \begin{subfigure}[b]{0.8\linewidth}
         \centering
         \includegraphics[width=\textwidth]{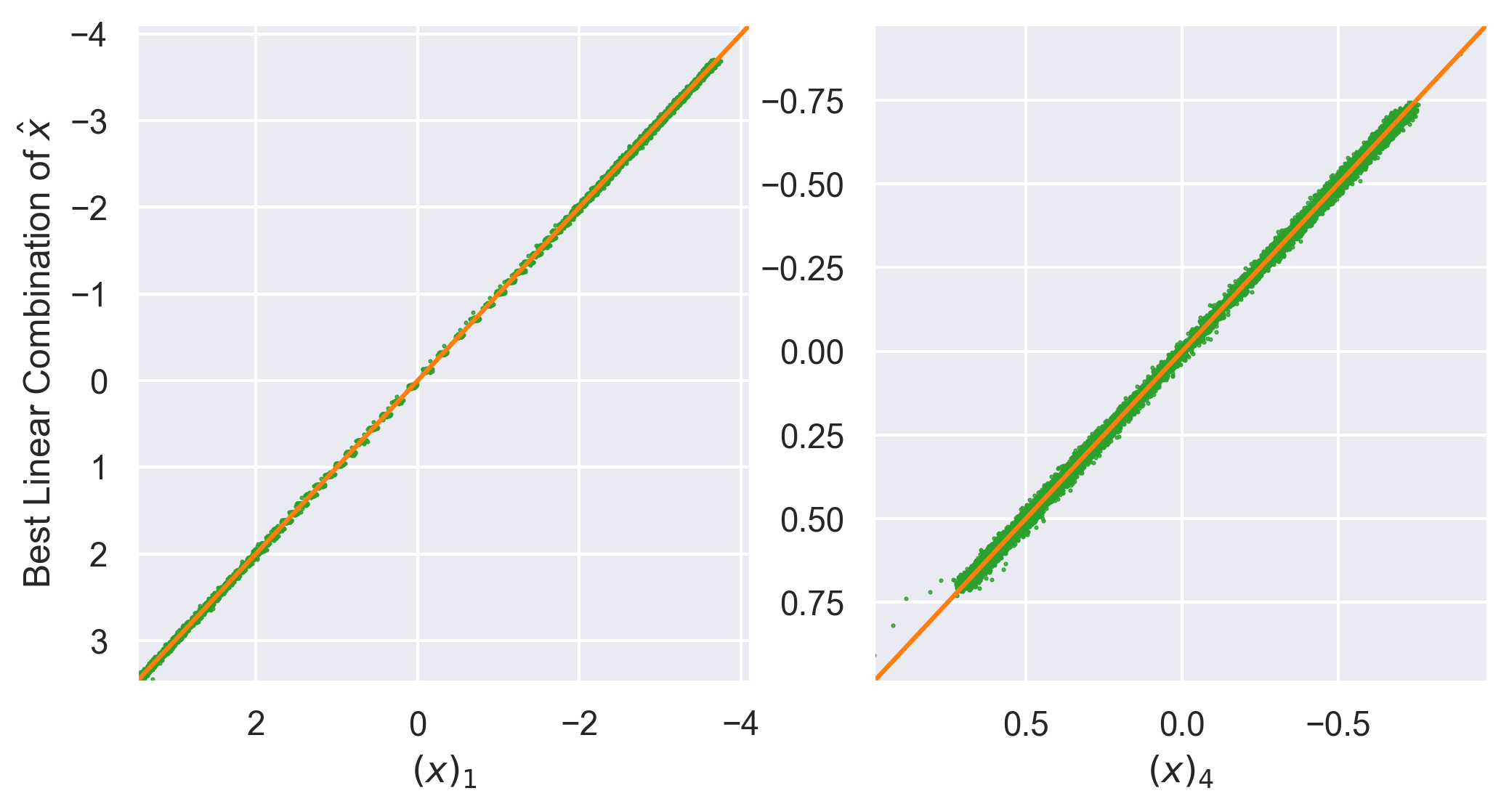}
         \caption{Correlation between the best linear combination of the internal states of the unstructured RNN and the true states.}
         \label{fig:linear:correlation:rra}
     \end{subfigure}
     \hfill
     \begin{subfigure}[b]{0.8\linewidth}
         \centering
         \includegraphics[width=\textwidth]{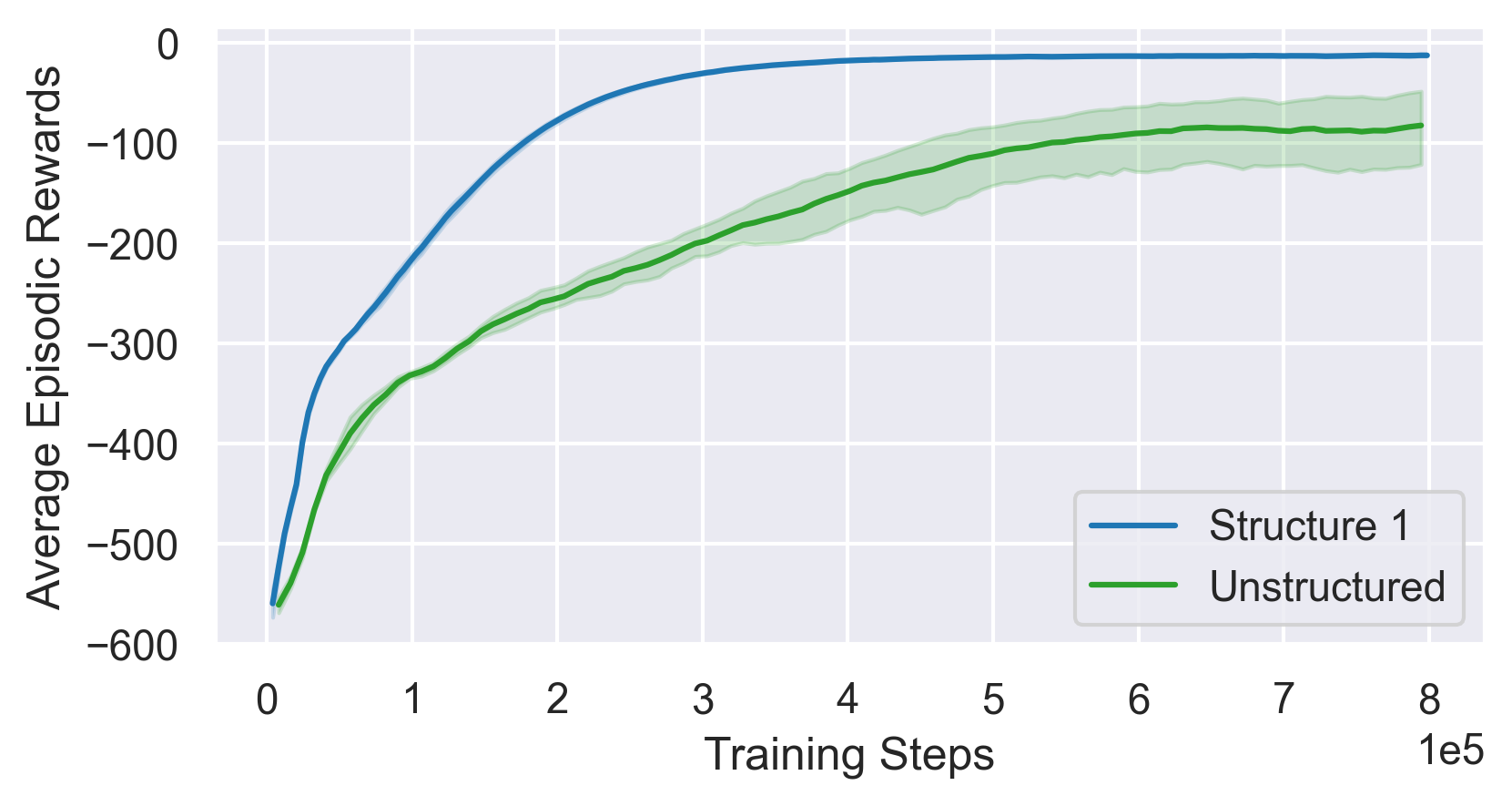}
         \caption{Training performance comparison.}
         \label{fig:linear:training}
     \end{subfigure}
     \hfill
     \begin{subfigure}[b]{0.8\linewidth}
         \centering
         \includegraphics[width=\textwidth]{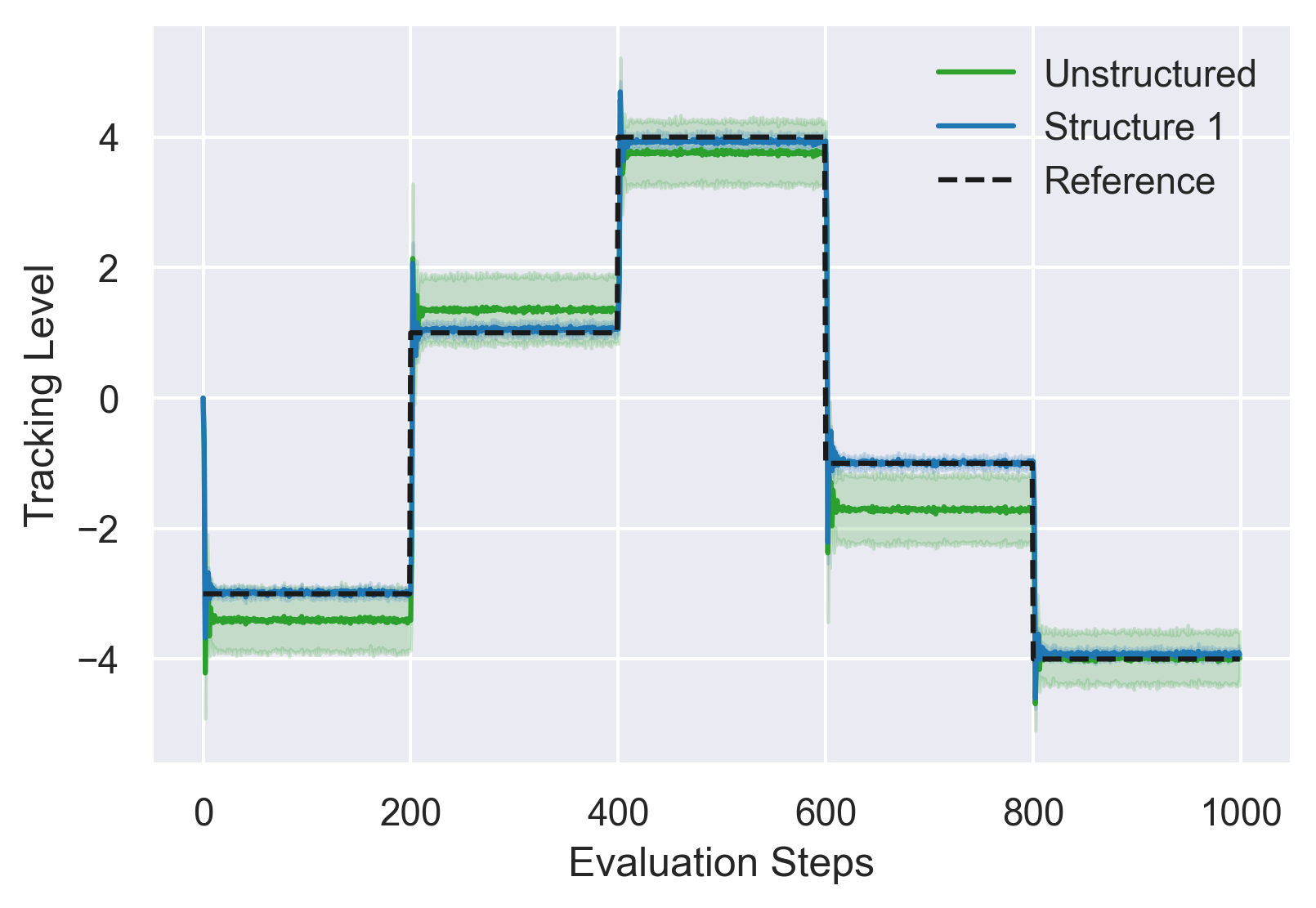}
         \caption{Evaluation comparison to track 5 different reference signals.}
         \label{fig:linear:eval}
     \end{subfigure}
          
     \caption{Performance comparison over 5 different training seeds for Linear 10D SISO system.}
     \label{fig:linear10D}
\end{figure}

Since the true order of the system is known, the dimension of the state  estimated by the observer in the structured network is set to 10, while the internal state of the unstructured RNN is given by the size of the network (64 in this case). To improve the training efficiency, both methods are combined with a prior P-controller as in \citep{pcontroller}.

To verify that the state $\hat{x}_t$ estimated by the structured observer depicted in Fig.~\ref{fig:obserferff-network-1} and the unstructured RNN are valid states for the system, the networks were first trained using PPO. The RL library Stable-Baselines3 by \citet{raffin2019stable} with modification was used for the implementation. 
Then data was collected by letting the trained controllers track 100 different reference levels. After this least squares was used to find the linear combination of $\hat{x}_t$ that corresponded the best with the true states $x_t$.
Fig.~\ref{fig:linear:correlation:roa} and \ref{fig:linear:correlation:rra} show the result for the 1st and 4th element of $x_t$, the results are similar for all other states. The \textit{orange} lines in the plots are the function $y=x$. 
The results show that the correlation is very high for the two methods, indicating that both learn a good state representation.

Fig~\ref{fig:linear:training} shows the average return the two methods experienced during training. It can be seen that the controller with the structured network learns a good controller significantly faster than the controller trained with the unstructured network. In Fig.~\ref{fig:linear:eval} the step response  performance of the controllers after $8 \times 10^5$ training steps is shown. It is clear that the structured network learns a more accurate controller with a smaller variance, as conjectured above. These results indeed indicate that the proposed structure can improve the performance of standard RL methods.

\section{Feedforward Structure  Performance}
\label{sec::vali:feedforward}

\begin{figure}[htbp!]
	\center
	\includegraphics[width=0.7\linewidth]{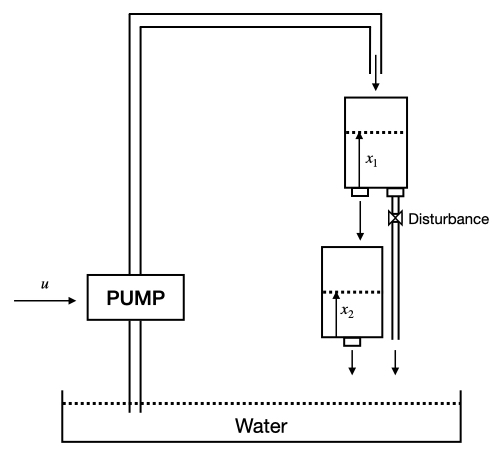}
	\caption{The cascaded double tank system.}
	\label{fig:watertank}
\end{figure}

In this section the proposed structures in Fig.~\ref{fig:obserferff-network-1}, referred to as ``Structure 1'', and Fig.~\ref{fig:obserferff-network-2}, referred to as ``Structure 2'', are compared with the use of unstructured RNNs as in \citep{heess2015memory-based} referred to as ``Unstructured''. 

Each structure was evaluated on a simulated water tank level control system implemented as the laboratory system depicted in Fig.~\ref{fig:watertank}. 
The control system consists of two tanks mounted on top of each other. The tank system dynamics are given by
\begin{align}
    \label{eq:env_watertank}
    \begin{split}
    \dot{x}_1(t) &= -\frac{a_1}{A_1}\sqrt{2gx_1(t)} + \frac{K_\text{pump}}{A_1}u(t) -\frac{d(t)}{A_1}\sqrt{2gx_1(t)},\\
    \dot{x}_2(t) &= \frac{a_1}{A_1}\sqrt{2gx_1(t)} - \frac{a_2}{A_2}\sqrt{2gx_2(t)}, \\
    y(t) &= x_2(t),
    \end{split}
\end{align}
where $x_i(t) \in \mathbb{R}^{+}$ is the level of the $i$th tank, $a_i$ is the area of the bottom outlets of the $i$th tank, $A_i$ is the area of the cross-section of the $i$th tank, $K_\text{pump}$ is the pump constant and $u(t) \in \mathbb{R}^{+}$ is the voltage applied to the pump.
The disturbance $d(t)$ considered here is another outlet in the top tank which area can be varied to allow the water into the water container.  Note that continuous time variables use the notation $y(t)$ while $y_t$ is used for the discrete time-step~$t$. 

For Structure 1 and Unstructured the measured disturbance $d_t$ is fed to the network together with the measured output $y_t$, but in Structure 2 a separate network is used the handle the disturbances. For both Structure 1 and Structure 2 the dimension of the complete estimated state  is~3, with two states for the tank levels and one for the disturbance. For the Unstructured alternative, the dimension of the internal state is given by the size of the network (64 in this case). 

The training and evaluation of the experiment were performed with the simulation of the tank system. In the simulation, the system was discretized using the Euler method, see e.g. \citet{glad2018controlbook}, with a sampling period of 2 seconds. The reward function at time step $t$ was given by 
\begin{align}
    r_t = -(y_t - \yref)^2.
    \label{eq:tank_reward}
\end{align}

During the training, the measurable disturbance amplitude was independently drawn with a uniform distribution in the interval $[0, a_1]$ every $50$ time step.
A prior P-controller was used to improve the exploration during training, as suggested by \citet{pcontroller}

\begin{figure}[t]
    \centering
    \begin{subfigure}[b]{\linewidth}
         \centering
         \includegraphics[width=\linewidth]{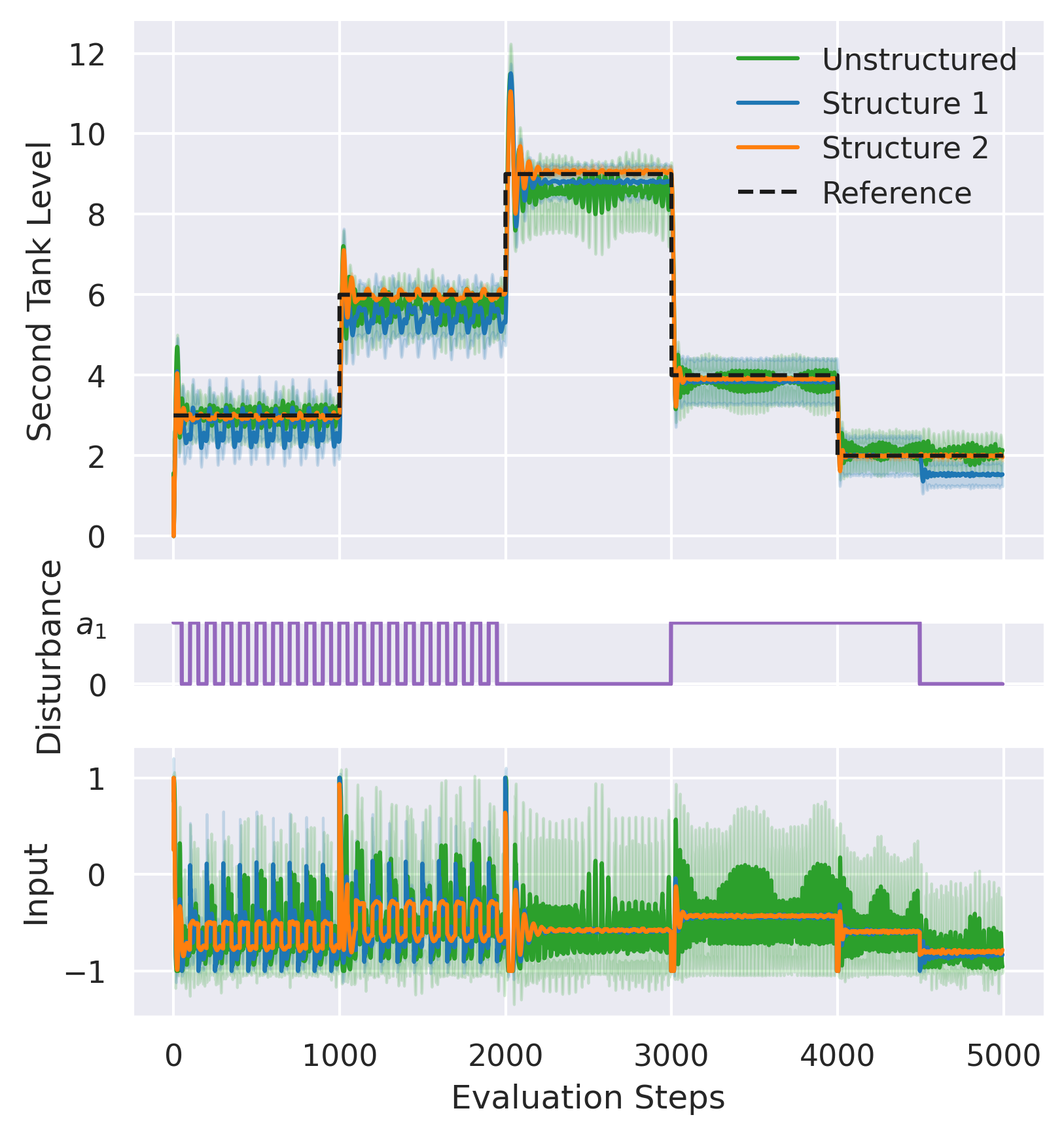}
         \caption{Combined feedback and feedforward control.}
         \label{fig:tank::d}
     \end{subfigure}

     \begin{subfigure}[b]{\linewidth}
         \centering
         \includegraphics[width=\linewidth]{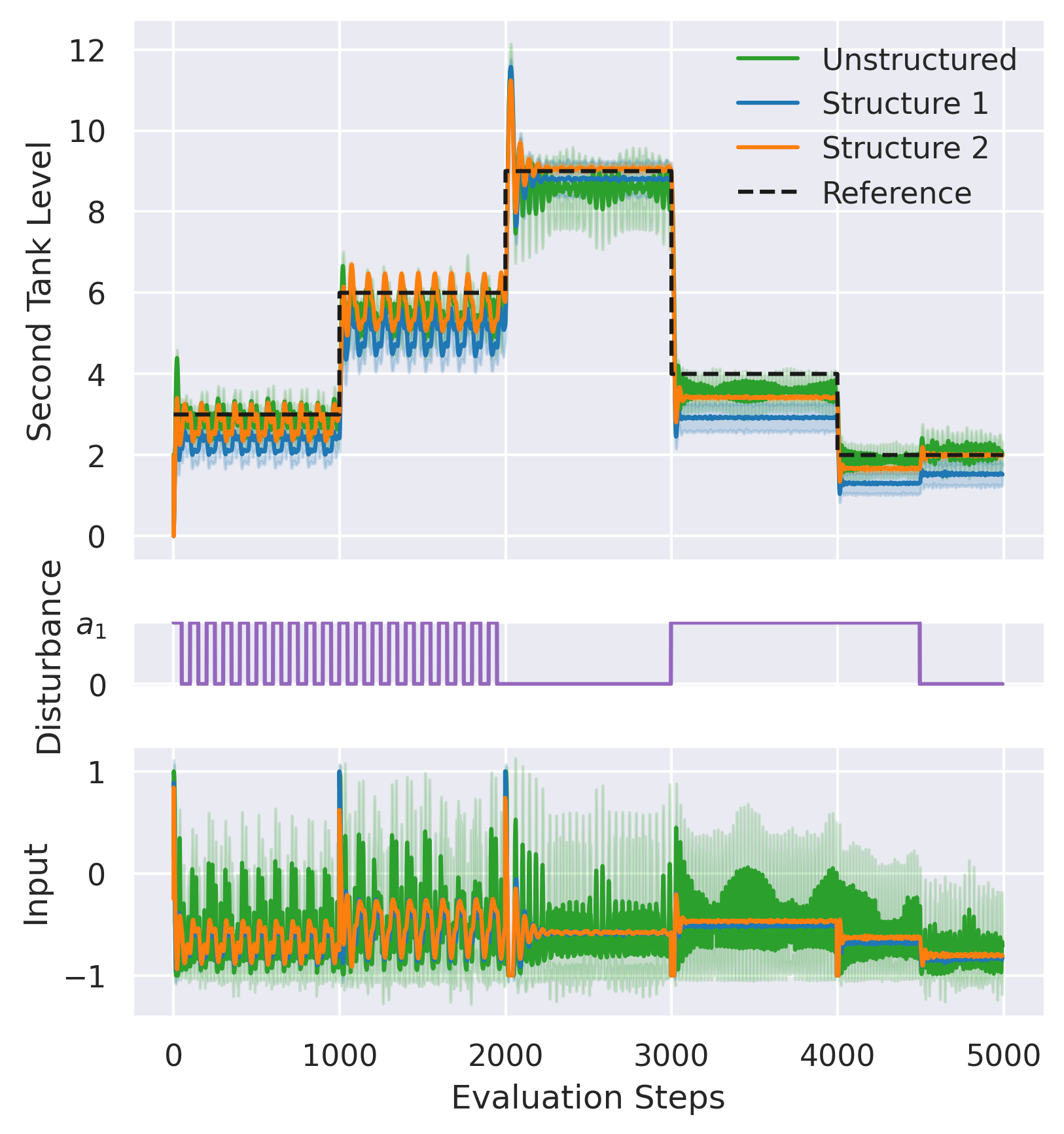}
         \caption{Control with the feedforward path disabled.}
         
         \label{fig:tank::invisible}
     \end{subfigure}
     \caption{Performance comparison when tracking 5 different levels with disturbances using four different training seeds.}
     \label{fig:tank}
\end{figure}

The performance of the RL based control systems is compared in Fig.~\ref{fig:tank} after training $8\times10^5$ time steps. 
Here Fig.~\ref{fig:tank::d} shows the case when combined feedback and feedforward is applied. It is clear that Structure 2 performs the best with far better disturbance rejection than Structure 1 and the Unstructured alternative, primarily for a switching disturbance but also for constant disturbances at the minimum and maximum disturbance levels. In fact  Structure 2 is the only one that gets the level of compensation right. The Unstructured alternative performs the worst and when comparing the input signals, the Unstructured alternative does not settle when the measurable disturbance is constant as do the structured control systems. The remaining oscillations make it questionable if the Unstructured controller has any feedforward effect at all, it rather appears to converge to a controller that induces a limit cycle, i.e. has a too high gain. The input signal of the controller based on Structure~2 also appears to have a smaller amplitude than that of Structure~1, which points to better robustness.

Fig.~\ref{fig:tank::invisible} shows the effect when no feedforward control is applied. That is, even though the disturbance is affecting the system, $d_t=0$ is fed into the RNNs.
The effect of the disturbance is significant, and the conclusion that the unstructured controller is ill-tuned is supported, referring to both the input and output signals. For constant disturbance levels, the feedback controller of Structure 2 achieves the best results. Finally, the high frequency measurable disturbance is rejected in Fig.~\ref{fig:tank::d}  while not in Fig.~\ref{fig:tank::invisible}. This illustrates that the overall controller of Structure 2 uses low bandwidth feedback together with high bandwidth feedforward, as it should according to classical control theory. This reasoning also shows that it is possible to use the structure to \emph{understand} the behavior of the controller.

\section{Conclusions} 
\label{sec::conclusion}
Two new observer-feedback-feedforward neural network structures were proposed to improve the training efficiency and to reject a measurable disturbance when nonlinear dynamic systems with partial observations are to be controlled. The effectiveness of the proposed models and methods was verified with two numerical examples. The first example showed that the RNNs act as observers and do learn to estimate a valid state of the system. The second example considered the addition of feedforward and verified that the proposed model structure with two separated observers together with the combined feedback and feedforward controller neural networks  performed the best, and provided excellent disturbance rejecting feedforward control. 

It is therefore concluded that the new structured networks allow for a significant reduction of  the orders of the observers and controllers  which makes the RL algorithm converge faster. Furthermore, the addition of a feedforward controller enables very efficient rejection of a significant disturbance.
The concluding recommendation is therefore to structure the neural networks as much as the available prior information allows when RL based adaptive nonlinear control is applied.

\bibliography{observerff}

\begin{thebibliography}{18}
\providecommand{\natexlab}[1]{#1}
\providecommand{\url}[1]{\texttt{#1}}
\providecommand{\urlprefix}{URL }
\expandafter\ifx\csname urlstyle\endcsname\relax
  \providecommand{\doi}[1]{doi:\discretionary{}{}{}#1}\else
  \providecommand{\doi}{doi:\discretionary{}{}{}\begingroup
  \urlstyle{rm}\Url}\fi

\bibitem[{{\AA}str{\"o}m and Wittenmark(2008)}]{aastrom2013adaptive_control}
{\AA}str{\"o}m, K.J. and Wittenmark, B. (2008).
\newblock \emph{Adaptive Control}.
\newblock Dover Publications.

\bibitem[{Elman(1990)}]{elman1990finding}
Elman, J.L. (1990).
\newblock Finding structure in time.
\newblock \emph{Cognitive science}, 14(2), 179--211.

\bibitem[{Glad and Ljung(2018)}]{glad2018controlbook}
Glad, T. and Ljung, L. (2018).
\newblock \emph{Control theory}.
\newblock CRC Press.

\bibitem[{Ha and Schmidhuber(2018)}]{ha2018recurrent_modelbased}
Ha, D. and Schmidhuber, J. (2018).
\newblock Recurrent world models facilitate policy evolution.
\newblock \emph{Advances in neural information processing systems}, 31.

\bibitem[{Hausknecht and Stone(2015)}]{hausknecht2015dqn_rnn_modelfree}
Hausknecht, M. and Stone, P. (2015).
\newblock Deep recurrent q-learning for partially observable mdps.
\newblock In \emph{2015 AAAI Fall Symposium Series}.

\bibitem[{Heess et~al.(2015)Heess, Hunt, Lillicrap, and
  Silver}]{heess2015memory-based}
Heess, N., Hunt, J.J., Lillicrap, T.P., and Silver, D. (2015).
\newblock Memory-based control with recurrent neural networks.
\newblock \emph{arXiv preprint arXiv:1512.04455}.

\bibitem[{Kaelbling et~al.(1998)Kaelbling, Littman, and
  Cassandra}]{kaelbling1998planning}
Kaelbling, L.P., Littman, M.L., and Cassandra, A.R. (1998).
\newblock Planning and acting in partially observable stochastic domains.
\newblock \emph{Artificial intelligence}, 101(1-2), 99--134.

\bibitem[{Meng et~al.(2021)Meng, Gorbet, and Kuli{\'c}}]{meng2021memory}
Meng, L., Gorbet, R., and Kuli{\'c}, D. (2021).
\newblock Memory-based deep reinforcement learning for pomdps.
\newblock In \emph{2021 IEEE/RSJ International Conference on Intelligent Robots
  and Systems (IROS)}, 5619--5626. IEEE.

\bibitem[{Mnih et~al.(2015)Mnih, Kavukcuoglu, Silver, Rusu, Veness, Bellemare,
  Graves, Riedmiller, Fidjeland, Ostrovski et~al.}]{mnih2015dqn}
Mnih, V., Kavukcuoglu, K., Silver, D., Rusu, A.A., Veness, J., Bellemare, M.G.,
  Graves, A., Riedmiller, M., Fidjeland, A.K., Ostrovski, G., et~al. (2015).
\newblock Human-level control through deep reinforcement learning.
\newblock \emph{Nature}, 518(7540), 529--533.

\bibitem[{Plappert et~al.(2018)}]{plappert2018multigoal}
Plappert, M. et~al. (2018).
\newblock Multi-goal reinforcement learning: Challenging robotics environments
  and request for research.
\newblock \emph{arXiv:1802.09464}.
\newblock \urlprefix\url{https://arxiv.org/pdf/1802.09464}.

\bibitem[{Raffin et~al.(2019)Raffin, Hill, Ernestus, Gleave, Kanervisto, and
  Dormann}]{raffin2019stable}
Raffin, A., Hill, A., Ernestus, M., Gleave, A., Kanervisto, A., and Dormann, N.
  (2019).
\newblock Stable baselines3.

\bibitem[{Schaul et~al.(2015)Schaul, Horgan, Gregor, and
  Silver}]{schaul2015universal-value-functions}
Schaul, T., Horgan, D., Gregor, K., and Silver, D. (2015).
\newblock Universal value function approximators.
\newblock In \emph{International conference on machine learning}, 1312--1320.
  PMLR.

\bibitem[{Schulman et~al.(2017)Schulman, Wolski, Dhariwal, Radford, and
  Klimov}]{schulman2017proximal}
Schulman, J., Wolski, F., Dhariwal, P., Radford, A., and Klimov, O. (2017).
\newblock Proximal policy optimization algorithms.
\newblock \emph{arXiv:1707.06347}.

\bibitem[{S{\"o}derstr{\"o}m(2002)}]{soderstrom2002stochastic_system}
S{\"o}derstr{\"o}m, T. (2002).
\newblock \emph{Discrete-time stochastic systems: estimation and control}.
\newblock Springer-Verlag.

\bibitem[{Sternad and S{\"o}derstr{\"o}m(1988)}]{sternad1988lqg}
Sternad, M. and S{\"o}derstr{\"o}m, T. (1988).
\newblock L{Q}{G}-optimal feedforward regulators.
\newblock \emph{Automatica}, 24(4), 557--561.

\bibitem[{Wigren(2017)}]{wigren2017loop}
Wigren, T. (2017).
\newblock Loop-shaping feedback and feedforward control for networked systems
  with saturation and delay.
\newblock \emph{Asian Journal of Control}, 19(4), 1329--1349.

\bibitem[{Zhang et~al.(2019)Zhang, Vikram, Smith, Abbeel, Johnson, and
  Levine}]{zhang2019solar}
Zhang, M., Vikram, S., Smith, L., Abbeel, P., Johnson, M., and Levine, S.
  (2019).
\newblock Solar: Deep structured representations for model-based reinforcement
  learning.
\newblock In \emph{International Conference on Machine Learning}, 7444--7453.
  PMLR.

\bibitem[{Zhang et~al.(2023)Zhang, Mattsson, and Torbjörn}]{pcontroller}
Zhang, R., Mattsson, P., and Torbjörn, W. (2023).
\newblock Aiding reinforcement learning for set point control.
\newblock In \emph{22nd IFAC World Congress}.
\newblock (Accepted for presentation).

\end{thebibliography}

\appendix
\end{document}